\documentclass[12pt,preprint]{aastex}

\slugcomment{To appear in ApJ Letters}

\usepackage{natbib}
\shorttitle{The progenitor of the FUor-type young eruptive star 2MASS
  J06593158$-$0405277}
\shortauthors{K\'osp\'al et al.}

\begin{document}

\title{The progenitor of the FUor-type young eruptive star 2MASS
  J06593158$-$0405277\thanks{Based on observations made with the
    Herschel Space Observatory. Herschel is an ESA space observatory
    with science instruments provided by European-led Principal
    Investigator consortia and with important participation from
    NASA.}}

\author{\'A. K\'osp\'al\altaffilmark{1},
  P. \'Abrah\'am\altaffilmark{1}, A. Mo\'or\altaffilmark{1}, M. Haas\altaffilmark{2},
  R. Chini\altaffilmark{2,3} and M. Hackstein\altaffilmark{2}}

\altaffiltext{1}{Konkoly Observatory, Research Centre for Astronomy and Earth
  Sciences, Hungarian Academy of Sciences, PO Box 67, H-1525 Budapest,
  Hungary}

\altaffiltext{2}{Astronomisches Institut, Ruhr-Universit\"at Bochum,
  Universit\"atsstra\ss{}e 150, D-44801 Bochum, Germany}

\altaffiltext{3}{Instituto de Astronom\'{i}a, Universidad Cat\'{o}lica
  del Norte, Antofagasta, Chile}

\begin{abstract}
Only a dozen confirmed FU Orionis-type young outbursting stars (FUors)
are known today; this explains the interest in the recent FUor
eruption of 2MASS~J06593158$-$0405277. Its outburst and expected
decline will be subject to numerous studies in the future. Almost
equally important for the understanding of the eruption mechanism,
however, is the physical characterization of the FUor's
precursor. Here we analyze unpublished archival data and summarize --
and partly revise -- all relevant photometry from optical to
submillimeter wavelengths. Our analysis implies that the FUor is
possibly associated with eight T\,Tauri star candidates and a strong
Class\,0 source. Adopting a distance of 450\,pc for the FUor, we
derive a quiescent bolometric luminosity and temperature of $L_{\rm
  bol}$ = 4.8\,L$_{\odot}$ and $T_{\rm bol}$ = 1190\,K , typical for
young Class\,II sources. The central star has a temperature of $T_{\rm
  eff}$ = 4000\,K, a mass of 0.75\,M$_{\odot}$, and an age of about
6$\times$10$^5$\,yr. The SED implies a circumstellar mass of 0.01 --
0.06\,M$_{\odot}$, and the system is surrounded by a faint infrared
nebulosity. Our results provide an almost complete picture of a FUor
progenitor, supporting the interpretation of future post-outburst
studies.

\end{abstract}



\section{Introduction}

Nascent Sun-like stars are surrounded by circumstellar material, which
builds up the growing protostar as matter is accreted through a
circumstellar disk. The accretion rate varies in time: the protostar's
normal accretion at a low rate
($\sim$10$^{-8}$\,M$_{\odot}$\,yr$^{-1}$) is occasionally interspersed
by brief episodes of highly enhanced accretion \citep{kenyon1990}. FU
Orionis-type variables (FUors) are the visible examples of episodic
accretion. During their episodic ``outbursts'', accretion rate from
the circumstellar disk onto the star increases by several orders of
magnitude, leading to a 5--6 mag optical brightening \citep{hk96,
  audard2014}. The first FUors which defined the class ({\it
  ``classical FUors''}) are in outburst for decades
\citep{herbig1977}. As a consequence, information on their progenitors
is almost non-existing. The most common assumption is that they are
Class\,II, or Class\,I/Class\,II transitional objects with massive
disks and (in many cases) extended envelopes. Recent results, however,
demonstrate that deeply embedded Class\,I objects (e.g., V1647~Ori,
\citealt{abraham2004b}; OO\,Ser, \citealt{kospal2007}) and Class\,II
objects with very small disk masses and no envelopes (e.g., HBC 722,
\citealt{miller2011}) could also show FUor-like eruptions. Another
class of young eruptive stars, called EXors after the prototype
EX\,Lup, exhibit typically smaller, shorter, repetitive outbursts
\citep{herbig1977,herbig2008}.

On November 30, 2014, \citet{maehara2014} reported the discovery of a
new outbursting star, 2MASS~J06593158$-$0405277. Optical spectra of
the object obtained by \citet{maehara2014} and by
\citet{hillenbrand2014}, and infrared spectra published by
\citet{reipurth2015} and by \citet{pyo2015} are very similar to those
of the classical FUors. Our group has been monitoring the area around
this source with a robotic telescope at Bochum observatory near Cerro
Armazones since 2010. The optical light curves show that the object
started a slow rising at the end of 2013, and quickly brightened by
about 3\,mag both in $r$ and $i$ compared to the quiescent level,
reaching peak brightness at the end of October 2014
\citep{hackstein2014}.

Thanks to the object's proximity to the Galactic plane, its position
was covered with a number of all-sky or Galactic plane surveys,
providing information on the system in its quiescent phase. Thus,
2MASS~J06593158$-$40405277 offers a unique opportunity to study the
progenitor of a classical FUor. Here, we collect and analyze archival
optical, infrared and submillimeter observations of the new
outbursting star (hereafter, ``the FUor''), and discuss the structure
of its circumstellar and interstellar environment. Our study will
serve as a basis for future work on the object in outburst, including
possible changes in the circumstellar matter induced by the eruption.


\section{Optical/infrared/submillimeter data}

\paragraph{Optical data.}

The FUor is visible in the digitized Palomar Observatory Sky Survey
images. However, its magnitudes in the USNO-B1.0 catalog
\citep{monet2003} probably suffer from source confusion due to a
close-by star 5$\farcs$7 to the north (2MASS~J06593168$-$0405224,
hereafter, ``N component''). We repeated the photometry using a small,
3$''$ radius aperture. For calibration, we selected all those nearby
stars within 10$'$ whose instrumental magnitudes differ less than
$\pm$0.1\,mag from that of our target. Plotting their USNO-B1.0
magnitudes as a function of the instrumental magnitude, the
distribution of data points could be fitted with a linear
relationship. We used this linear fit to determine the magnitude of
the FUor (Table~\ref{tab:data}). Due to its faintness and proximity to
the FUor, no reasonable photometry could be obtained for the N
component.

\paragraph{Near-infrared data.}
JHK$_S$ magnitudes for the FUor in quiescent phase are available in
the 2MASS catalog \citep{cutri2003}. The source was also covered in
the DENIS \citep{epchtein1998} and UKIDSS \citep{lawrence2007}
near-infrared sky surveys. For the sake of homogeneous calibration, we
performed aperture photometry on the DENIS and UKIDSS images, using a
3$''$ radius aperture, and adopting nearby sources from the 2MASS
catalog as photometric standards in the $JHK_S$ bands. In the DENIS
$I$ band, we used the magnitudes of the comparison stars from the
DENIS catalog. The resulting near-infrared photometry is listed in
Table~\ref{tab:data}. Additionally, we obtained photometry for a
number of fainter sources visible in the UKIDSS images in the vicinity
of the FUor, whose colors will be analyzed in
Sec.~\ref{sec:envi}. These values are listed in Table~\ref{tab:data2}.

\paragraph{Mid- and far-infrared data.}

The area around the FUor was observed by the Infrared Array Camera
(IRAC, \citealt{fazio2004}) on-board the Spitzer Space Telescope
\citep{werner2004} at 3.6 and 4.5$\,\mu$m. We took photometry for the
FUor from the GLIMPSE360 Catalog \citep{churchwell2009} and obtained
psf-photometry for the N component using the idl-based tool {\sc
  Starfinder} \citep{diolaiti2000}. The FUor was also covered by the
Midcourse Space Experiment (MSX, \citealt{egan2003}), resulting a
8.28$\,\mu$m flux in the MSX6C Point Source Catalog. The Wide-field
Infrared Survey Explorer (WISE, \citealt{wright2010}) observed our
source in all four bands on April 1--2, 2010 in the cryogenic mission,
and in the two shorter wavelength bands on October 10--11, 2010, as
part of the post-cryo mission. At 3.35 and 4.60$\,\mu$m we used data
from the ALLWISE Multiepoch Photometry Table and averaged the April and
October data points separately. At 11.56 and 22.09$\,\mu$m, we used the
ALLWISE Source Catalog, that combines all data from the WISE
mission. The FUor was also covered by the AKARI mission
\citep{akari}. It is present in the AKARI/IRC All-Sky Survey Point
Source Catalogue, with photometry at 9 and 18$\,\mu$m.

\paragraph{Far-infrared and submillimeter data.}

The area around the FUor was observed with the Herschel Space
Observatory \citep{pilbratt2010} in the PACS/SPIRE parallel mode
\citep{poglitsch2010,griffin2010}. We processed the PACS 70 and
160\,{\micron} data with the Herschel Interactive Processing
Environment v.13 \citep[HIPE,][]{ott2010} using the standard script
optimized for mapping observations. We used the recently developed
gyro correction to reduce the pointing jitter. The final 70 and
160$\,\mu$m maps were made with the {\sc photProject} task. At
70$\,\mu$m, two point sources are detectable: one corresponding to the
FUor, while the other source is located 5$\farcs$9 to the southeast
(hereafter, ``SE component''). We used 3$''$ radius apertures to
measure the flux ratio of these two targets, then measured the total
flux in a large 22$''$ radius aperture, and divided it between the two
targets according to their flux ratio. At 160$\,\mu$m, the two sources
are not well resolved any more, but the centroid of the image is
within 1$\farcs$2 of the 70$\,\mu$m position of the SE component,
indicating that at this wavelength, the SE source dominates the
emission. At this wavelength, photometry was obtained in a 22$''$
radius aperture. The fluxes, whose uncertainties include the 7\%
overall calibration uncertainty of the PACS photometer, are presented
in Table~\ref{tab:data}. To derive photometry in the SPIRE 250, 350,
and 500$\,\mu$m bands, we used the {\sc Timeline Fitter} task of HIPE
\citep{bendo2013} to fit circular Gaussians to the baseline-subtracted
timeline data. Here again, the centroid of the source corresponds to
the SE component, and not to the FUor. The resulting fluxes, whose
uncertainties include the 5.5\% overall calibration uncertainty of the
SPIRE photometer, are presented in Table~\ref{tab:data}. The AKARI/FIS
All-Sky Survey Point Source Catalogue contains a source which is
closer to the SE component than to the FUor. While the photometry at
65$\,\mu$m probably suffers from confusion of these two targets, the
fluxes at longer wavelengths can be safely attributed to the SE
component.


\section{Results and analysis}

\subsection{The broader environment}
\label{sec:envi}

\begin{figure}
\includegraphics[angle=90,scale=.65]{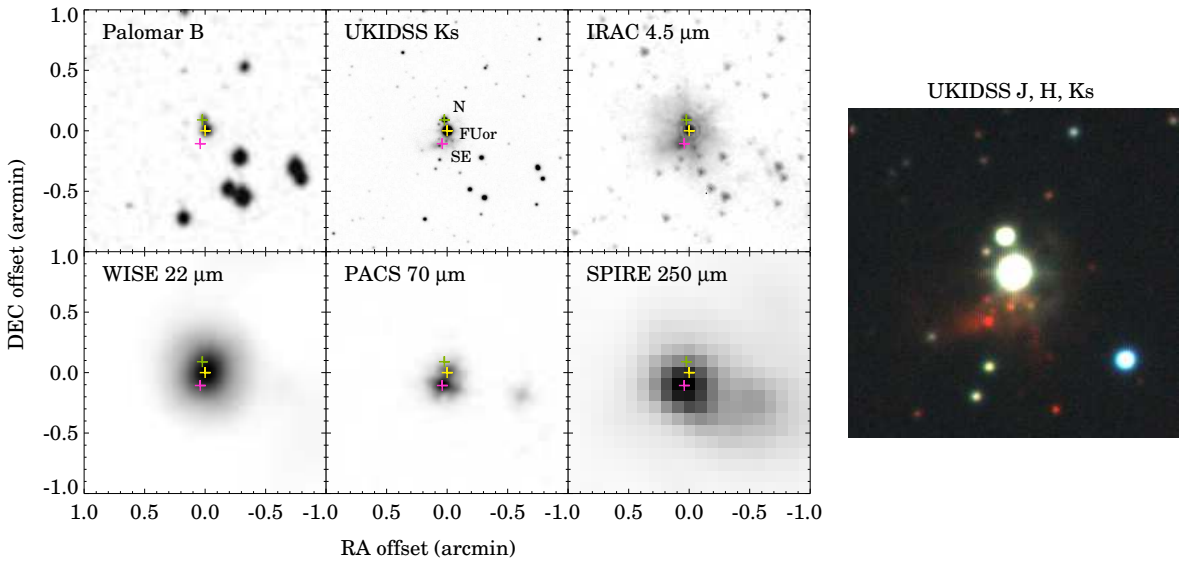}
\caption{2MASS~J06593158$-$0405277 and its surroundings at different
  wavelengths. Yellow plus sign marks the optical stellar position of
  the FUor, magenta plus sign marks the SE component, and green plus
  marks the N component. The zoom-in on the right shows a
  color-composite image of an area of 50$''\,{\times}\,$50$''$
  centered on the FUor using $J$ as blue, $H$ as green, and $K_S$ as
  red.
\label{fig:6panels}}
\end{figure}

Fig.~\ref{fig:6panels} displays the region around the FUor at
different wavelengths revealing a complex environment. The FUor itself
is well visible between the optical and 70$\,\mu$m. The N component is
located at a separation of 5$\farcs$7, PA of 15$^{\circ}$, clearly
discernible until 4.5$\,\mu$m, and possibly present even in the WISE
bands. The SE component is located at a distance of 5$\farcs$9, PA of
157$^{\circ}$. This source is first visible at 70$\,\mu$m, and we
could not unambiguously link it to any point sources at shorter
wavelengths. At 70$\,\mu$m it is already brighter than the FUor and
clearly dominates the emission at longer wavelengths.

The deep UKIDSS images reveal a faint nebulosity around the FUor,
especially apparent to the south but visible to the west as well
(Fig.~\ref{fig:6panels}, right). Its characteristic size is $\approx$
15$''$. This extended emission is also discernible in the IRAC
images. The nebulosity, especially its southern part, is much redder
than the FUor itself. If the origin of the nebulosity is light from
the FUor scattered by the circumstellar matter, then the nebulosity
should be bluer than the light source, unless it is reddened by some
dust along the line-of-sight. We speculate that the extinction needed
for this may be provided by the massive envelope of the SE component.

In order to find out whether the FUor is an isolated young star or
there are other pre-main sequence objects in its vicinity, we obtained
photometry on the UKIDSS images for several point sources within 1$'$
of the FUor. By plotting the objects on a $J-H$ vs.~$H-K_S$
color-color diagram (Fig.~\ref{fig:tcd}, left), we found that most of
them are on the main sequence, or they are reddened main sequence
stars with extinction up to $A_V = 6$\,mag. In addition to the FUor,
we found eight stars around the T\,Tauri locus \citep{meyer1997}, one
being the N component, while the other seven are marked in
Table~\ref{tab:data2}. This indicates that they are young stellar
object candidates, thus the FUor is probably not isolated.

\subsection{Variability in quiescence}

In Table~\ref{tab:data} we collected all available pre-outburst
photometry for the FUor. In the 0.4--5$\,\mu$m range, multi-epoch data
is available, making it possible to search for variability in the
quiescent phase. Significant flux changes up to 0.7\,mag are revealed
when comparing the two different red Palomar points, and also the
infrared Palomar and DENIS $I$-band points. In the $J$, $H$, and $K_S$
bands and in the 3--5$\,\mu$m range, the object was constant within
0.15\,mag. This indicates that the object has a well-defined quiescent
state, with accretion rate changes probably much smaller than what
caused the recent outburst.

\subsection{Colors and spectral energy distribution}
\label{sec:colors}

Based on the photometry in Table~\ref{tab:data}, we plotted the
pre-outburst spectral energy distribution (SED) of the FUor in
Figure~\ref{fig:sed} with dots, the N component with squares, and the
SE component with asterisks. We only used points which can be
unambiguously attributed to either of these three sources. Thus, the
AKARI/FIS point at 65$\,\mu$m, where the FUor and the SE source are
probably blended together, is omitted.

The SED of the FUor indicates significant infrared emission from the
circumstellar matter. On the $J-H$ vs.~$H-K_S$ color-color diagram
(Fig.~\ref{fig:tcd}, left), the object falls onto the T\,Tauri locus
\citep{meyer1997}, suggesting that it is a regular T\,Tauri star with
negligible reddening ($A_V <$ 0.5\,mag). Indeed, its
optical--near-infrared SED (up to the $H$ band) can be reasonably well
fitted using a Kurucz model of a late K-type or early M-type
photosphere with effective temperatures between 3750\,K and 4250\,K
(Fig.~\ref{fig:sed}). Excess emission compared to the stellar
photosphere appears first in the $K_S$ band. Until about 12$\,\mu$m,
the SED decreases like ${\nu}F_{\nu} \sim \lambda^{-0.5}$, typical for
moderately flared disks \citep{kenyon1987}. After about 12$\,\mu$m,
the SED starts rising again, suggesting the presence of an outer
envelope.

The N component has a SED shape similar to the FUor, indicating that
it is probably a T\,Tauri star as well, but with lower temperature and
mass. Indeed, it falls onto the T\,Tauri locus in the near-infrared
color-color diagram (see also Sec.~\ref{sec:envi} and
Fig.~\ref{fig:tcd}, left). The SE component is probably a deeply
embedded Class\,0 object, invisible in the optical, and peaking at
around 160$\,\mu$m. Its submillimeter spectral slope is ${\nu}F_{\nu}
\sim \lambda^{-2.5}$.

\subsection{Distance}

The FUor is situated in the Galactic plane ($l$ = 217.5, $b$ =
$-$0.1). The Herschel images of this area suggest that it sits on top
of a long, elongated filament corresponding to the LDN\,1650 molecular
cloud (also known as S\,287). CO observations by, e.g.,
\citet{kim2004} suggest a velocity of $v_{\rm
  LSR}$ = 26--28\,km\,s$^{-1}$ for the molecular gas, which gives a
kinematic distance of 2.3\,kpc. However, supposing that the FUor is a
low-mass pre-main sequence star, as suggested by its
optical--near-infrared SED (Sect.~\ref{sec:colors}), its apparent
brightness contradicts this relatively large distance.

The pre-main sequence evolutionary tracks of \citet{siess2000} predict
that the absolute $J$ magnitude of a star in the 3750--4250\,K
temperature range should be between 2 and 4\,mag for ages between
5$\times$10$^5$ and 2$\times$10$^6$\,yr (Fig.~\ref{fig:tcd},
right). The apparent $J$ = 11\,mag would then correspond to a distance
range of 250--630\,pc for the FUor. If the object is located at
2.3\,kpc, its absolute $J$ magnitude would be $-$0.8\,mag, typical for
much higher temperature young objects. The observed SED, however,
peaks at near-infrared wavelengths, and could be consistent with a
higher temperature photosphere only if significantly reddened. For
instance, a $T_{\rm eff}$ = 6000\,K photosphere with $A_V$ = 2.3\,mag
reddening would adequately fit the SED until the $J$ band. Thus, due
to the well-known degeneracy between the stellar temperature and the
interstellar extinction, the central star can either be an unreddened
low-mass M-type young star at a distance of about 450\,pc (a
representative value between 250 and 630\,pc), or a moderately
reddened intermediate-mass F-type young star at a distance of
2.3\,kpc.

Without a pre-outburst optical spectrum that could be used for stellar
classification, it is very difficult to decide whether the FUor is an
unreddened M-star close-by, or a reddened F-star far away. There is,
however, extra information that we can use to decide. We scaled a
blackbody function to the 70$\,\mu$m flux of the FUor, extrapolated to
850$\,\mu$m, and used this flux prediction to estimate the total mass
of the circumstellar matter. We adopted a gas-to-dust mass ratio of
100. Assuming 55\,K or 100\,K for the dust temperature, we obtained
0.06\,M$_{\odot}$ or 0.01\,M$_{\odot}$, respectively, for a distance
of 450\,pc, and 1.6\,M$_{\odot}$ or 0.28\,M$_{\odot}$ for a distance
of 2.3\,kpc. A similar estimate can be done for the SE component using
the 500$\,\mu$m SPIRE flux. Here we used 50\,K or 20\,K, and obtained
0.31\,M$_{\odot}$ or 1.3\,M$_{\odot}$, respectively, for 450\,pc, and
8.2\,M$_{\odot}$ or 34\,M$_{\odot}$ for 2.3\,kpc. In the Galactic
plane there is no guarantee that the FUor and the SE source is at the
same distance. However, the fact that there is at least 2 (or possible
a few more) T\,Tauri stars and an embedded submillimeter source all
within 1$'$ strongly suggests that they are associated to each
other. In this case, the larger distance would give an unrealistically
large total mass for the SE component. Therefore, we suggest that the
young stellar objects in this area, including new outbursting star,
are not associated with LDN\,1650, but located much closer to us,
probably around $d$ = 450\,pc. We will use this distance value in our
subsequent discussion.


\section{Discussion}

\begin{figure}[h!]
\epsscale{0.85}
\plotone{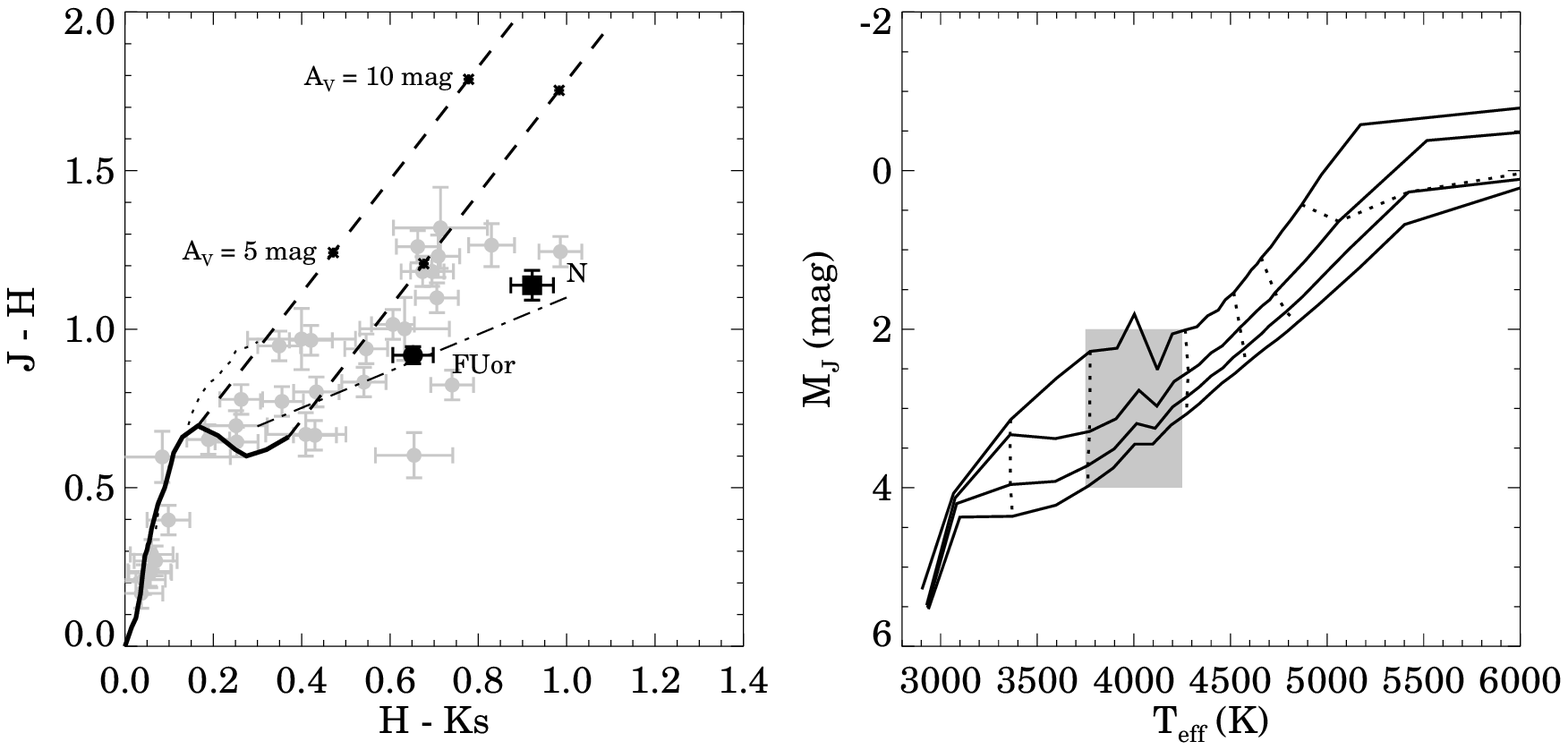}
\caption{Left: Near-infrared color-color diagram. The main-sequence is
  marked by a solid line, the giant branch with dotted line
  \citep{koornneef1983}, the reddening path with dashed lines
  \citep{cardelli1989}, and the T\,Tauri locus with dash-dotted line
  \citep{meyer1997}. The black dot marks the FUor
  (2MASS~J06593158$-$0405277), while the black square marks the N
  component (2MASS~J06593168$-$0405224). Right: PMS evolutionary
  tracks for 0.5, 1.0, 1.5, and 2.0 Myr (from top to bottom) from
  \citet{siess2000}. The dotted lines mark stellar masses of 0.3, 0.5,
  1.0, 1.5, 2.0, and 3.0\,M$_{\odot}$ (from left to right). The
  temperature range allowed by assuming that the central star is a
  late K or early M-type star with negligible reddening and the
  corresponding range in absolute $J$ magnitudes is marked with a gray
  rectangle.
\label{fig:tcd}}
\end{figure}

\begin{figure}
\epsscale{0.80}
\plotone{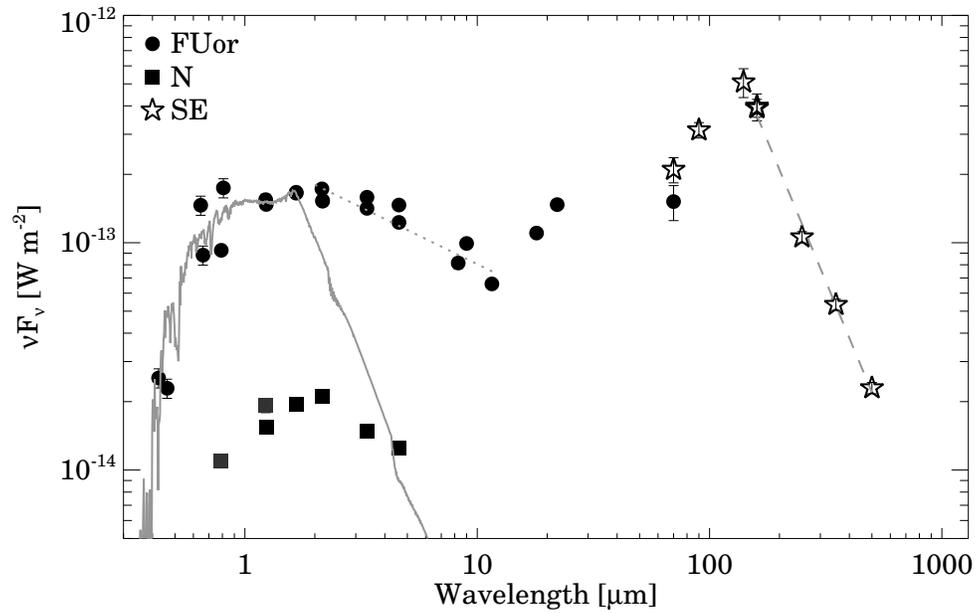}
\caption{Pre-outburst SED of the FUor (2MASS~J06593158$-$0405277) with
  black dots, the northern component (2MASS~J06593168$-$0405224), and
  the SE (submillimeter) component. See Table~\ref{tab:data} for the
  source of the data points. The gray curve is a 4000\,K Kurucz model,
  the dotted line is a power law with a slope of $-0.5$, and the
  dashed line is a power law with a slope of $-2.5$. \label{fig:sed}}
\end{figure}

In the following, we discuss the physical nature of the FUor in
quiescence. The central star is a low-mass young star, with effective
temperature of about $T_{\rm eff}$ = 4000\,K $\pm$ 250\,K
(Sec.~\ref{sec:colors}). The Siess models in Fig.~\ref{fig:tcd}
predict a stellar mass of $M_*$ = 0.75\,$M_{\odot} \pm$
0.25\,$M_{\odot}$. By integrating the SED in Fig.~\ref{fig:sed}, we
obtained a stellar luminosity of $L_*$ = 1.2\,$L_{\odot}$ and a total
bolometric luminosity is $L_{\rm bol}$ = 4.8\,$L_{\odot}$. Following
\citet{chen1995}, we determined the bolometric temperature of the
object, which is a distance-independent quantity. The resulting
$T_{\rm bol}$ = 1190\,K indicates that the object is Class\,II (their
Fig.~6), and its age is approximately 6$\times$10$^5$\,yr (their
Fig.~4).

The circumstellar structure of the FUor resembles the canonical
picture of classical FUors \citep{hk96}, where a young T\,Tauri star
is surrounded by a moderately flared circumstellar disk and an
envelope. In our case, the total circumstellar mass is on the order of
0.01--0.06\,M$_{\odot}$, falling into the range of typical T\,Tauri
disks \citep{beckwith1990}, and is somewhat lower than envelope masses
measured by \citet{sw2001} and by \citet{kospal2011} for FUors. The
SED shows a conspicuous break at 12$\,\mu$m, which may indicate an
inner hole in the envelope. Additionally, the envelope must have a
cavity as well, through which we have an unobscured view of the star
and the inner disk.

The comparison of the SED with other classical FUors is not
straightforward, because for them, only outburst SEDs are known
\citep{abraham2004}. For this reason, we compare the SED in
Fig.~\ref{fig:sed} to the quiescent SED of the young eruptive star
V1647\,Ori dereddened by $A_V$=13\,mag \citep[Fig.~3
  in][]{abraham2004b}. The two SEDs exhibit similar shapes, and it is
remarkable that even their absolute flux levels agree within a factor
of 3 in the whole 1--70$\,\mu$m range, strengthening our distance
estimate for the FUor.  From an evolutionary point of view,
\citet{quanz2007} and \citet{green2006} proposed that FUors are young
stars in transition from the Class\,I to Class\,II. The transition
process is associated with the gradual dispersion of the envelope due
to the repetitive outbursts. In this picture, our results on the
progenitor of the classical FUor candidate 2MASS~J06593158$-$0405277
indicate that it is an evolved FUor, approaching the evolutionary
phase of a disk-only T\,Tauri star.


\acknowledgments

This work was supported by the Momentum grant of the MTA CSFK
Lend\"ulet Disk Research Group, and the Hungarian Research Fund OTKA
grant K101393. A.~M.~acknowledges support from the Bolyai Research
Fellowship of the Hungarian Academy of Sciences.

{\it Facilities:} \facility{Spitzer}, \facility{Herschel},
\facility{WISE}, \facility{AKARI}.



\begin{deluxetable}{ccccccc}
\tabletypesize{\scriptsize}
\rotate
\tablecaption{Photometry of the FUor 2MASS~J06593158$-$0405277 and nearby objects\label{tab:data}}
\tablewidth{0pt}
\tablehead{
\colhead{Instrument/Catalog} & \colhead{Filter/Wavelength} &
\colhead{Date} & \colhead{Magnitude/Flux} & \colhead{Magnitude/Flux} &
\colhead{Magnitude/Flux}  & \colhead{Reference}\\
                   &       \colhead{($\mu$m)}     &        &
  \colhead{FUor}               &    \colhead{N component}         &
  \colhead{SE component}    &      
}
\startdata
Palomar        & B1 / 0.425 & 1953-01-17	 & 15.20 $\pm$ 0.1 mag    & \dots   	   	   & \dots	     & 1 \\
Palomar        & R1 / 0.645 & 1953-01-17	 & 12.47 $\pm$ 0.1 mag    & \dots   	   	   & \dots	     & 1 \\
Palomar        & B2 / 0.463 & 1983-01-17	 & 15.18 $\pm$ 0.1 mag    & \dots   	   	   & \dots	     & 1 \\
Palomar        & R2 / 0.660 & 1989-03-05	 & 12.97 $\pm$ 0.1 mag    & \dots   	   	   & \dots	     & 1 \\
Palomar        & I  / 0.808 & 1982-01-22	 & 11.82 $\pm$ 0.1 mag    & \dots   	   	   & \dots	     & 1 \\
2MASS          & J  / 1.235 & 1998-11-02	 & 11.049 $\pm$ 0.030 mag & 13.500 $\pm$ 0.054 mag & \dots	     & 2 \\
2MASS          & H  / 1.662 & 1998-11-02	 & 10.122 $\pm$ 0.036 mag & 12.445 $\pm$ 0.068 mag & \dots	     & 2 \\
2MASS          & Ks / 2.159 & 1998-11-02	 & 9.452  $\pm$ 0.027 mag & 11.611 $\pm$ 0.029 mag & \dots	     & 2 \\
DENIS          & I  / 0.791 & 1997-01-11	 & 12.525 $\pm$ 0.020 mag & 14.842 $\pm$ 0.027 mag & \dots	     & 1 \\
DENIS          & J  / 1.228 & 1997-01-11	 & 11.003 $\pm$ 0.060 mag & 13.266 $\pm$ 0.075 mag & \dots	     & 1 \\
DENIS          & Ks / 2.145 & 1997-01-11	 & 9.329  $\pm$ 0.050 mag & 11.488 $\pm$ 0.088 mag                & \dots	     & 1 \\
UKIDSS         & J  / 1.235 & 2010-01-01	 & 11.030 $\pm$ 0.022 mag & 13.688 $\pm$ 0.031 mag & \dots	     & 1 \\
UKIDSS         & H  / 1.662 & 2010-01-01	 & 10.112 $\pm$ 0.017 mag & 12.550 $\pm$ 0.035 mag & \dots	     & 1 \\
UKIDSS         & Ks / 2.159 & 2010-01-01	 & 9.460 $\pm$ 0.043 mag  & 11.628 $\pm$ 0.033 mag & \dots	     & 1 \\
UKIDSS         & Ks / 2.159 & 2006-11-30	 & 9.382 $\pm$ 0.058 mag  & 11.690 $\pm$ 0.037 mag & \dots	     & 1 \\
Spitzer/IRAC   &      3.6   & 2011-06-05         & 8.121 $\pm$ 0.045 mag  & 10.570 $\pm$ 0.070 mag & \dots	     & 3 \\
Spitzer/IRAC   &      4.5   & 2011-06-05         & 7.448 $\pm$ 0.026 mag  & 9.925 $\pm$	0.030 mag  & \dots	     & 3 \\
MSX            & A  / 8.28  & 1996-05 -- 1997-01 & 5.987 $\pm$ 0.059 mag  & \dots    		   & \dots	     & 4 \\
WISE           & W1 / 3.35  & 2010-04-01/02	 & 8.063 $\pm$ 0.023 mag  & \dots    		   & \dots	     & 5 \\
WISE           & W1 / 3.35  & 2010-10-10/11	 & 8.106 $\pm$ 0.023 mag  & \dots    		   & \dots	     & 5 \\
WISE           & W2 / 4.60  & 2010-04-01/02	 & 7.186 $\pm$ 0.022 mag  & \dots    		   & \dots	     & 5 \\
WISE           & W2 / 4.60  & 2010-10-10/11	 & 7.232 $\pm$ 0.020 mag  & \dots    		   & \dots	     & 5 \\
WISE           & W3 / 11.56 & 2010-04-01/02	 & 5.240 $\pm$ 0.016 mag  & \dots    		   & \dots	     & 6 \\
WISE           & W4 / 22.09 & 2010-04-01/02	 & 2.219 $\pm$ 0.018 mag  & \dots    		   & \dots	     & 6 \\
AKARI/IRC      &      9     & 2006-05 -- 2006-11 & 5.690 $\pm$ 0.022 mag  & \dots    		   & \dots	     & 7 \\
AKARI/IRC      &      18    & 2006-05 -- 2006-11 & 3.146 $\pm$ 0.049 mag  & \dots    		   & \dots	     & 7 \\

AKARI/FIS      &      65    & 2006-05 -- 2006-11 & \dots 		  & \dots    		   & 6.723 $\pm$ 1.826 Jy  & 8 \\
AKARI/FIS      &      90    & 2006-05 -- 2006-11 & \dots 		  & \dots   		   & 9.396 $\pm$ 0.734 Jy  & 8 \\
AKARI/FIS      &      140   & 2006-05 -- 2006-11 & \dots 		  & \dots    		   & 23.751 $\pm$ 3.448 Jy & 8 \\
AKARI/FIS      &      160   & 2006-05 -- 2006-11 & \dots 		  & \dots    		   & 21.203 $\pm$ 2.879 Jy & 8 \\
Herschel/PACS  &      70    & 2011-05-08	 & 3.54 $\pm$ 0.62 Jy	  & \dots    		   & 4.90 $\pm$ 0.62 Jy    & 1 \\
Herschel/PACS  &      160   & 2011-05-08	 & \dots		  & \dots    		   & 20.9 $\pm$ 1.89 Jy & 1 \\
Herschel/SPIRE &      250   & 2011-05-08	 & \dots		  & \dots    		   & 8.83 $\pm$ 0.50 Jy & 1 \\
Herschel/SPIRE &      350   & 2011-05-08	 & \dots		  & \dots    		   & 6.24 $\pm$ 0.35 Jy & 1 \\
Herschel/SPIRE &      500   & 2011-05-08	 & \dots		  & \dots    		   & 3.83 $\pm$ 0.22 Jy & 1 \\
\enddata
\tablecomments{References: 1 -- this work; 2 -- 2MASS All-Sky Catalog
  of Point Sources \citep{cutri2003}; 3 -- GLIMPSE360 Catalog
  \citep{churchwell2009}; 4 -- MSX6C Infrared Point Source Catalog
    \citep{egan2003}; 5 -- ALLWISE Multiepoch Photometry Table
    \citep{cutri2014}; 6 -- ALLWISE Source Catalog \citep{cutri2014};
    7 -- AKARI/IRC All-Sky Survey Point Source Catalogue
    \citep{akariirc}; 8 -- AKARI/FIS All-Sky Survey Point Source
    Catalogue \citep{akarifis}}
\end{deluxetable}

\begin{deluxetable}{cccccc}
\tabletypesize{\scriptsize}
\tablecaption{UKIDSS photometry for sources within 1$'$ of the FUor 2MASS~J06593158$-$0405277\label{tab:data2}}
\tablewidth{0pt}
\tablehead{
 & \colhead{RA$_{J2000}$} & \colhead{DEC$_{J2000}$} & \colhead{$J$ mag} & \colhead{$H$ mag} & \colhead{$K_S$ mag}}
\startdata
  & 06:59:28.47 & $-$04:05:51.84 & 13.25 $\pm$ 0.03 & 12.96 $\pm$ 0.04 & 12.90 $\pm$ 0.03 \\
  & 06:59:28.63 & $-$04:05:46.55 & 12.64 $\pm$ 0.03 & 12.41 $\pm$ 0.04 & 12.35 $\pm$ 0.03 \\
T & 06:59:28.67 & $-$04:06:04.54 & 16.10 $\pm$ 0.03 & 15.00 $\pm$ 0.04 & 14.30 $\pm$ 0.03 \\
T & 06:59:28.90 & $-$04:04:50.50 & 18.44 $\pm$ 0.06 & 17.17 $\pm$ 0.04 & 16.34 $\pm$ 0.03 \\
  & 06:59:28.95 & $-$04:05:01.30 & 17.23 $\pm$ 0.03 & 16.57 $\pm$ 0.04 & 16.14 $\pm$ 0.03 \\
T & 06:59:29.10 & $-$04:05:11.11 & 17.89 $\pm$ 0.03 & 16.65 $\pm$ 0.04 & 15.66 $\pm$ 0.03 \\
  & 06:59:29.58 & $-$04:05:34.60 & 16.83 $\pm$ 0.03 & 16.19 $\pm$ 0.04 & 15.93 $\pm$ 0.03 \\
  & 06:59:30.00 & $-$04:05:05.06 & 17.83 $\pm$ 0.06 & 16.60 $\pm$ 0.04 & 15.89 $\pm$ 0.03 \\
  & 06:59:30.11 & $-$04:06:14.84 & 18.28 $\pm$ 0.05 & 17.61 $\pm$ 0.05 & 17.20 $\pm$ 0.08 \\
  & 06:59:30.31 & $-$04:04:56.04 & 15.04 $\pm$ 0.03 & 14.84 $\pm$ 0.04 & 14.86 $\pm$ 0.03 \\
  & 06:59:30.36 & $-$04:04:57.08 & 14.82 $\pm$ 0.03 & 14.43 $\pm$ 0.04 & 14.33 $\pm$ 0.03 \\
  & 06:59:30.39 & $-$04:06:01.22 & 12.22 $\pm$ 0.03 & 12.06 $\pm$ 0.04 & 12.02 $\pm$ 0.03 \\
  & 06:59:30.49 & $-$04:05:41.42 & 12.87 $\pm$ 0.03 & 12.66 $\pm$ 0.04 & 12.61 $\pm$ 0.03 \\
  & 06:59:30.51 & $-$04:06:09.42 & 16.43 $\pm$ 0.03 & 15.49 $\pm$ 0.04 & 14.95 $\pm$ 0.03 \\
  & 06:59:30.87 & $-$04:05:57.09 & 13.13 $\pm$ 0.03 & 12.86 $\pm$ 0.04 & 12.79 $\pm$ 0.03 \\
  & 06:59:30.94 & $-$04:06:22.94 & 17.25 $\pm$ 0.03 & 16.48 $\pm$ 0.04 & 16.13 $\pm$ 0.03 \\
  & 06:59:31.01 & $-$04:05:06.78 & 16.39 $\pm$ 0.03 & 15.73 $\pm$ 0.04 & 15.55 $\pm$ 0.03 \\
  & 06:59:31.40 & $-$04:06:20.01 & 17.65 $\pm$ 0.03 & 16.68 $\pm$ 0.04 & 16.26 $\pm$ 0.03 \\
  & 06:59:31.87 & $-$04:05:42.41 & 16.28 $\pm$ 0.03 & 15.09 $\pm$ 0.04 & 14.42 $\pm$ 0.03 \\
T & 06:59:31.89 & $-$04:05:24.93 & 16.53 $\pm$ 0.03 & 15.71 $\pm$ 0.04 & 14.97 $\pm$ 0.03 \\
  & 06:59:32.00 & $-$04:05:46.88 & 16.31 $\pm$ 0.03 & 15.36 $\pm$ 0.04 & 15.02 $\pm$ 0.03 \\
  & 06:59:32.36 & $-$04:06:11.75 & 13.75 $\pm$ 0.03 & 13.51 $\pm$ 0.04 & 13.45 $\pm$ 0.03 \\
T & 06:59:32.45 & $-$04:05:37.84 & 17.21 $\pm$ 0.03 & 16.38 $\pm$ 0.04 & 15.83 $\pm$ 0.04 \\
T & 06:59:32.47 & $-$04:04:49.12 & 18.89 $\pm$ 0.08 & 17.89 $\pm$ 0.06 & 17.25 $\pm$ 0.08 \\
T & 06:59:32.71 & $-$04:05:11.40 & 18.32 $\pm$ 0.05 & 17.72 $\pm$ 0.05 & 17.06 $\pm$ 0.07 \\
  & 06:59:32.85 & $-$04:05:12.88 & 18.49 $\pm$ 0.06 & 17.90 $\pm$ 0.06 & 17.81 $\pm$ 0.14 \\
  & 06:59:32.97 & $-$04:04:59.10 & 19.12 $\pm$ 0.11 & 17.81 $\pm$ 0.07 & 17.09 $\pm$ 0.08 \\
  & 06:59:33.08 & $-$04:04:49.46 & 15.57 $\pm$ 0.03 & 14.55 $\pm$ 0.04 & 13.94 $\pm$ 0.03 \\
  & 06:59:33.26 & $-$04:05:06.37 & 17.71 $\pm$ 0.03 & 16.53 $\pm$ 0.04 & 15.84 $\pm$ 0.03 \\
  & 06:59:33.61 & $-$04:06:05.09 & 18.90 $\pm$ 0.08 & 17.93 $\pm$ 0.06 & 17.53 $\pm$ 0.11 \\
  & 06:59:33.74 & $-$04:05:32.00 & 18.06 $\pm$ 0.04 & 16.80 $\pm$ 0.04 & 16.14 $\pm$ 0.03 \\
  & 06:59:34.15 & $-$04:05:21.60 & 17.89 $\pm$ 0.03 & 17.19 $\pm$ 0.04 & 16.94 $\pm$ 0.06 \\
  & 06:59:34.75 & $-$04:06:12.32 & 17.47 $\pm$ 0.03 & 16.67 $\pm$ 0.04 & 16.24 $\pm$ 0.04 \\
  & 06:59:34.76 & $-$04:04:56.32 & 17.01 $\pm$ 0.03 & 16.23 $\pm$ 0.04 & 15.97 $\pm$ 0.03 \\
\enddata
\tablecomments{T marks T Tauri star candidates.}
\end{deluxetable}

\end{document}